\documentclass[preprint]{aastex}

\newcommand{\etal}{{et al.\ }}

\newcommand{\be}{\begin{equation}}
\newcommand{\ee}{\end{equation}}
\newcommand{\bea}{\begin{eqnarray}}
\newcommand{\eea}{\end{eqnarray}}

\shorttitle{Hot Jupiters and the Roche limit}
\shortauthors{Ford \& Rasio}

\begin{document}
\title{On the Relation Between Hot Jupiters and the Roche Limit}
\author{Eric B.\ Ford$^1$, Frederic A.\ Rasio$^2$}
\affil{$^1$Astronomy Department, UC Berkeley, 601 Campbell Hall,
        Berkeley, CA 94709}
\email{eford@astro.berkeley.edu}
\affil{$^2$Physics and Astronomy, Northwestern University, 2145  
Sheridan Rd,
        Evanston, IL 60208}
\email{rasio@northwestern.edu}
\begin{abstract}
Many of the known extrasolar planets are ``hot Jupiters,''
giant planets with orbital periods of just a few days.
We use the observed distribution of hot Jupiters to constrain the
location of its inner edge in the mass--period diagram.  If we
assume a slope corresponding to the classical Roche
limit, then we find that the edge corresponds to a separation close  
to {\it twice\/} the Roche
limit, as expected if the planets started on highly
eccentric orbits that were later circularized. In contrast,
any migration scenario would predict an inner edge right at the
Roche limit, which applies to planets approaching on nearly circular  
orbits.
However, the current sample of
hot Jupiters is not sufficient to provide a precise constraint  
simultaneously
on both the location and slope of the inner edge.
\end{abstract}


\section{Introduction}
Early discoveries of hot Jupiters hinted at
a pile-up near a 3-day period, but recent transit surveys and more  
sensitive
radial velocity observations have discovered planets with even shorter
periods. The data now suggest that the inner limit for hot
Jupiters is not defined by an orbital period, but rather by a tidal
limit, which depends on both the separation and the planet-star
mass ratio (Fig.\ 1).  This would arise naturally if the inner
edge were related to the Roche limit, the critical distance within  
which a
planet would start losing mass (Faber et al.\ 2005).  The Roche limit  
separation,
$a_R$, is given by $ R_P = 0.462\, a_R \,\mu^{1/3}$,
where $R_P$ is the radius of the planet, and $\mu=m/M_*$ is the  
planet-star
mass ratio.

The many formation scenarios proposed for hot Jupiters can be divided  
into two broad
categories.  The first involves slow migration on quasi-circular  
orbits, perhaps due to
interaction with a
gaseous disk or planetesimal scattering (Murray et al.\ 1998; Trilling \etal~1998).
This would result in an inner edge precisely at the Roche limit.
The second category invokes tidal circularization of highly eccentric
orbits with very small pericenter distances, following
planet-planet scattering (Rasio \& Ford 1996; Weidenschilling \&  
Marzari 1996;
Ford et al.\ 2001; Papaloizou \& Terquem 2001; Marzari \&  
Weidenschilling
2002), secular perturbations from a wide binary companion (Holman et  
al.\ 1997;
Wu \& Murray 2003), or tidal capture of free-floating planets
(Gaudi 2003).  These would result in a limiting separation of {\it
twice\/} the Roche limit, assuming that circularization can take  
place without
significant mass loss from the planet\footnote{This is very easy to show:
consider a  planet on an initially
eccentric orbit,  with initial  eccentricity  $e$  and
pericenter  distance $r_p$.  Circularizing  this   orbit  under  ideal
conditions leads to dissipation of energy but conservation of mass and
angular momentum. Simply equating the angular momentum of the initial
and final orbits gives a final circularized radius
$a  = r_p (1+e) \simeq 2 r_p$  for $e \simeq 1$.}  (Faber et al.\ 2005; 
Gu et al.\ 2003; Rasio et al.\ 1996).

\section{Statistical Analysis}

To constrain rigorously the
distribution of hot Jupiters, we adopt a Bayesian
framework, where the model parameters are
treated as random variables to be constrained by the actual
observations.  To perform a Bayesian analysis it is
necessary to specify both the likelihood (the probability of making a
certain observation given a particular set of model parameters) and
the prior (the {\em a priori\/} probability distribution for the
model parameters).  Let us denote the model parameters by $\theta$ and
the data by $d$, so that their joint probability
distribution function (PDF) is given by
$
p(d, \theta) = p(\theta) p(d | \theta) = p(d) p(\theta | d).
$
Note how the joint PDF is expanded in two ways,
both expressed as the product of a marginalized PDF
and a conditional PDF.  The prior is
given by $p(\theta)$ and the likelihood by $p(d|\theta)$, while $p(d)$
is the {\em a priori\/} probability for observing the
values actually measured and $p(\theta|d)$ is the PDF
of primary interest: the {\em a posteriori\/} PDF
for the model parameters conditioned on the actual
observations.  The probability of the
observations $p(d)$ can be
obtained by marginalizing over the joint PDF and again
expanding the joint density as the product of the prior and the
likelihood.  This leads to Bayes' theorem, the primary tool for
Bayesian inference,
\be
p(\theta | d ) = \frac{p(d|\theta)p(\theta)}{p(d)} = \frac{p(d|\theta) 
p(\theta)}{\int \, d\theta p(d|\theta)p(\theta)}
\ee
Often the model parameters contain a quantity of
particular interest (the location of the inner cutoff for hot
Jupiters in our analysis) plus other ``nuisance parameters,'' which are
necessary to describe the observations (e.g., the fraction
of stars with hot Jupiters in our analysis).  Since Bayes' theorem
provides a real PDF for the model parameters, we
can simply marginalize over the nuisance parameters to calculate a
marginalized posterior PDF, which will be our basis
for making inferences about the location of the inner cutoff for hot
Jupiters.

\subsubsection{1-d Model}

We start by presenting a simple 1-d model for the
distribution of hot Jupiters.  The primary question we wish to
address is the location of the inner edge of the
distribution relative to the Roche
limit.  Therefore, we define $x \equiv a / a_R$, where $a$ is the
semimajor axis of the planet and $a_R$ is the Roche limit.  We assume
that the actual distribution of $x$ for hot Jupiters is given
by a truncated power law,
\be
p(x | \gamma, x_l, x_u) dx = x^\gamma \left(\frac{dx}{x}\right),  
\quad x_l < x < x_u,
\ee
and zero elsewhere.  Here $\gamma$ is the power-law index and $x_l$
and $x_u$ are the lower and upper limits for $x$.  The lower limit,
$x_l$, is the model parameter of primary interest, while $\gamma$ and
$x_u$ are nuisance parameters.  Therefore, our results are contained in
the marginalized posterior PDF for $x_l$.

For simplicity we restrict our analysis to the subset of
known extrasolar planets discovered by complete radial-velocity surveys,
extremely unlikely to contain any false positives.  To
obtain such a sample, we impose two constraints: $P\le P_{\max}$, where
$P_{\max}$ is a maximum orbital period, and $K\ge K_{\min}$, where
$K_{\min}$ is a minimum velocity semi-amplitude.  We use
$K_{\min}=30$m/s, following Cumming (2004).  We typically set $P_ 
{\max}=30\,$d, even though
radial-velocity surveys are likely to be complete for even longer
periods (provided $K\ge K_{\min}$).  This minimizes the chance of
introducing biases due to survey incompleteness or possible structure  
in the
observed distribution at larger periods.  By
considering only planets with orbital parameters such that radial- 
velocity
surveys are very nearly complete, our analysis does not
depend on the velocities of stars for which no planet has been
detected.  Note that our criteria for including a planet may
introduce a bias depending on the actual mass-period distribution.
We will address this with a 2-d model below.
Note that, in this paper, we exclude any planet
discovered via techniques other than radial velocities (e.g.,
transits), even if subsequent radial-velocity observations were
obtained to confirm the planet.

Initially, we make several simplifying assumptions to allow for a  
simple analytic
treatment.  We assume uniform priors
for each of the model parameters, $p(\gamma) \sim
U(\gamma_{\min},\gamma_{\max})$ and $p(x_l,x_u) \sim \mathrm{const}$,
provided $x_{ll} < x_l < x_u < x_{uu}$ and zero otherwise.  The lower
and upper limits are chosen
to be sufficiently far removed from regions of high likelihood that
these choices do not affect our results.
We assume that the orbital period ($P$), velocity semi-amplitude
($K$), semi-major axis ($a$), stellar mass ($M_*$), and planet mass
($m \sin i$) are known exactly based on the observations.

We begin by assuming that $\sin i=1$ (orbital plane seen nearly edge- 
on) for all systems
and that all planets have the same radius, $R_P$.
With these
assumptions, the posterior probability distribution is
\be
p(x_l, x_u, \gamma | x_1, ... x_n ) \sim \gamma^n (x_u^\gamma - x_l^ 
\gamma)^{-n}
\prod_{j=1}^n x_j^{\gamma-1},
\ee
provided that
$x_{ll} < x_l \le x_{(1)} \le  x_{(n)} \le x_u < x_{uu}$ and
$\gamma_{\min} < \gamma < \gamma_{\max}$.  Here
$n$ is the number of planets included
in the analysis, $x_{(1)}$ is the smallest value of $x$ among the
planets used in the analysis, and $x_{(n)}$ is the largest
value.  The normalization can be obtained by
integrating over all allowed values of $x_l$, $x_u$, and $\gamma$.
We show the marginal posterior distributions after
integrating over the nuisance parameters, $x_u$ and $\gamma$, in Fig.~2
(dotted line), assuming $R_P = 1.2\,R_J$.  The distribution has a sharp
cutoff at $x_{(1)}$ and a tail to lower values reflecting the chance
that $x_l < x_{(1)}$ due to the finite sample size.

Next, we adopt an isotropic distribution of inclinations
($\cos i\sim~U[-1,1]$), but we use the measured value
for radial-velocity planets where the orbital
inclination has been determined via transits.  We show the marginal
posterior distribution for $x_l$ in Fig.~2 (solid line).
The sharp cutoff at $x_{(1)}$ is replaced with a more gradual tail,
reflecting the chance that $\sin i<1$ for planets with the smallest
values of $x$.

Now consider the consequences of allowing for a distribution of
planetary radii.  For transiting planets we use a normal distribution
based on the published radius value and uncertainty. For
non-transiting planets, we assume
a normal distribution with standard deviation $\sigma_{R_P}$.  We  
show the resulting
marginalized posterior PDFs in Fig.~2.  Allowing for a
significant dispersion broadens the posterior distribution for $x_l$
and results in a slight shift to smaller values.
We have also explored the effects of varying the model parameter
$P_{\max}$ from $8\,$d to $60\,$d.  We find that this does not
produce any discernible difference in the posterior PDF for $x_l$.

Our results are sensitive to the choice of mean radius for the
non-transiting planets.  In Fig.~3 (upper) we show the posterior
PDFs for various mean radii, assuming $\sigma_{R_P}
= 0.1\,R_J$.  Since few planets have a known inclination, there is
a nearly perfect degeneracy between $R_P$ and $x_l$.  Even when
we include transiting planets, this degeneracy remains near perfect,
i.e., $p(x_l | R_p, x_1, ..., x_n ) \simeq p(x_l \cdot \frac{R'_P} 
{R_P} | R'_P, x_1, ..., x_n)$.
However, it remains extremely
unlikely that $x_l\simeq1$ for any reasonable planetary
radius.

\subsubsection{2-d Model}

We can improve our analysis by more properly considering the
joint PDF in orbital period and planet-star mass ratio, which we  
write as
\be
p(P, \mu | \alpha, \beta, P_{\min}, P_{\max}, \mu_{\min}, \mu_{\max},  
c ) \sim c P^\alpha \mu^\beta \frac{dP}{P} \frac{d\mu}{\mu},
\ee
provided $\mu_{\min} < \mu < \mu_{\max}$, $P < P_{\max}$, and $a(P,  
M_*) \ge x_l \cdot a_R(R_P, \mu)$.  Here $\alpha$ and $\beta$ are new  
power-law indices.  Again our
results are not sensitive to the nuisance
parameters, $\mu_{\min}$, $\mu_{\max}$, and $P_{\max}$.  For
definiteness, we fix their values at $P_{\max} = 30\,$d, $\mu_{\min}
= 3.3\times10^{-5}$, and $\mu_{\max} = 0.01$.  We take priors uniform in
$\tan^{-1}(\alpha)$ and $\tan^{-1}(\beta)$, as the density
$U[-\pi/2,\pi/2]$ corresponds to uniform prior density for the slope
of the power-law distribution on a log-log plot.  We take a prior
uniform in $\log c$, as is standard for scale parameters.  Our
calculation of the likelihood is similar to that of Tabachnik \&
Tremaine (2002), except we replaced their inner boundary of $P\ge
P_{\min}$ with our boundary $a(P, M_*) \ge x_l a_R(R_P,\mu)$.  The
necessary integrals can be performed analytically provided we
approximate $1+\mu\simeq1$.  For convenience we used
{\tt Maple} to obtain an analytic expression
for the integral of the likelihood over $P$ and $\mu$.  The remaining
integrals over $\alpha$, $\beta$, and $c$ were performed numerically
over a grid with $\sim10^{10}$ points.

By considering the joint mass-period PDF, we are
able to account for the bias previously introduced by imposing
$K\ge K_{\max}$.  Since we are considering only planets with
orbital parameters such that radial-velocity surveys are very nearly
complete, our results depend only on the number of surveyed stars for
which no planet has been discovered, and not the observed velocities
of these stars.  For the total number of stars in radial-velocity
surveys that are complete for $K\ge K_{\min}$ and
$P\le P_{\max}$ we estimate $N_* = 2000$.
We show the resulting marginalized posterior PDF for
$x_l$ in Fig.\ 3 (lower).

\subsubsection{The Shape of the Inner Cutoff}

The above results clearly demonstrate that the present
observations strongly favor an inner cutoff at the ideal
circularization distance rather than at the Roche limit, {\em  
assuming that
the inner edge follows the slope of the Roche limit\/}.  We have also  
performed
calculations treating this slope
as an unknown model parameter.  Unfortunately, the present  
observations are not
sufficient to constrain this parameter empirically, and the resulting
marginalized posterior PDF for $x_l$ still allows, but no longer
exclusively favors, $x_l \simeq 2$.

We could gain additional leverage by including planets in short-period
orbits down to lower $K_{\min}$.  Unfortunately, the incompleteness of
the radial-velocity surveys is likely to be significant for $K_{\min}
< 20\,$m/s.  Since the number, quality, and spacing of observations
varies widely among the stars in these surveys, it would be
necessary to calculate the probability for detecting planets as a
function of orbital period and velocity semi-amplitude for each
star.
Additionally, the planet mass-radius relation, which is extremely
flat for planets near $1\,M_J$, becomes important for much
lower-mass planets.  Therefore, we have not attempted to extend our
analysis to planets for which radial-velocity surveys are not
yet complete.  We expect that the recent improvements in measurement
precision will eventually extend their completeness to smaller
$K_{\min}$.

We have begun a preliminary investigation of the constraints obtained
by adding information from the OGLE transit survey (Udalski 2002).
Due to both signal-to-noise and aliasing issues, the OGLE survey does
not provide a complete sample of short-period planets for a
significant fraction of parameter space.  Therefore, it is necessary
to calculate the probability of detecting planets with various orbital
periods and radii.  We estimate these probabilities by taking the actual
observation times and uncertainties for
the 62 transit candidates from the 2002 OGLE observations of the
Carina field and applying the detection criteria from Pont et al.\ (2005).
While the transiting planets are consistent with our above findings,
they do not provide sufficient additional information to constrain the
slope of the inner edge.  We look forward to both radial-velocity
and transit surveys detecting additional lower-mass objects in
short-period orbits, so that we may eventually constrain the slope of
the cutoff empirically.

\section{Discussion}

The current distribution of hot Jupiters shows a cutoff that is a  
function of orbital period and
planet mass. Under the assumption that the slope of this cutoff  
follows the Roche limit,
our Bayesian analysis solidly rejects the hypothesis
that the cutoff occurs inside or at the present Roche limit. This is  
in constrast to
what would be expected if these planets had slowly migrated inwards
on  quasi-circular orbits and with radii close to the presently measured
values around $1.2\,R_{J}$.  If confirmed by future analyses of a more
extensive data set, this result would be highly significant, as it  
would eliminate
a broad class of popular migration scenarios for the formation of hot  
Jupiters.

Instead, our analysis shows that this cutoff
occurs at a distance nearly twice that of the Roche limit, as expected
if the planets had been circularized from a highly eccentric orbit.
These findings suggest that hot Jupiters may have formed via
planet-planet scattering (Rasio \& Ford 1996), tidal capture of
free floating planets (Gaudi 2003), or secular perturbations from a
highly inclined binary companion (Holman et al.\ 1997).  Regardless  
of the
exact mechanism,
our model would require that the hot Jupiters all started on highly  
eccentric
orbits and survived the strong tidal dissipation needed to  
circularize their
orbits. A few caveats are worth mentioning.
Our study addresses the statistical properties
of the population of hot-Jupiters  and does not attempt to advance the
state of  knowledge of any specific planet.
In particular, we adopt
average properties of an assumed distribution that is analogous 
to---and derived from---the presently known distribution of hot Jupiters,
but we do not consider or solve for the specific properties of any 
individual extrasolar planet.
Moreover, strongly non-random or non-gaussian effects would be poorly modeled with the 
technique developed here.

An alternative explanation is that the planets migrated inwards at
an early time and arrived at their Roche limit on a quasi-circular orbit
when their radii were still $\ge 2\,R_{J}$ (Burrows et al.\ 2000). The
dissipation process causing the migration must then have stopped  
immediately
afterwards to avoid further decay of the orbit as the planets  
continued to
cool and contract. We find this scenario unattractive, especially since
there is no natural explanation for the factor of 2 in this case.

Yet another alternative is that short-period giant planets are destroyed
by some process {\em before\/} they reach the Roche limit.  HST
observations of HD 209458 indicate absorption by matter presently
beyond the Roche lobe of the planet and have been interpreted as  
evidence
for a wind leaving the planet and powered by stellar irradiation  
(Vidal-Madjar et
al.\ 2003, 2004).  Further theoretical work will help determine under
what conditions these processes can cause significant mass loss
(e.g., Hubbard et al.\ 2005) and whether complete destruction could  
occur rather
suddenly when the orbital radius decreases below $\sim 2 a_{R}$.

Future planet discoveries will either tighten the
constraints on the model parameters or provide evidence for the
existence of planets definitely closer than twice the Roche limit.
Additionally, future discoveries of
transiting hot-Jupiters around young stars could help discriminate  
between
the above alternatives.
Moreover, new detections of lower-mass planets with very short  
periods could help better
constrain the shape of the inner cutoff as a function of mass.  In the
future, an improved statistical analysis could also include such low- 
mass planets,
where surveys are not yet complete.

\acknowledgments{We thank E.\ Chiang, T.\ Loredo, N.\ Murray,
R.\ Murray-Clay, J.\ Papaloizou, F.\ Pont, and an anonymous  
referee for helpful comments.
This research was supported by NSF grants AST-0206182 and AST-0507727
at Northwestern University and by a Miller Research Fellowship to EBF  
at UC Berkeley.
}
\begin{figure}[ht]
\plotone{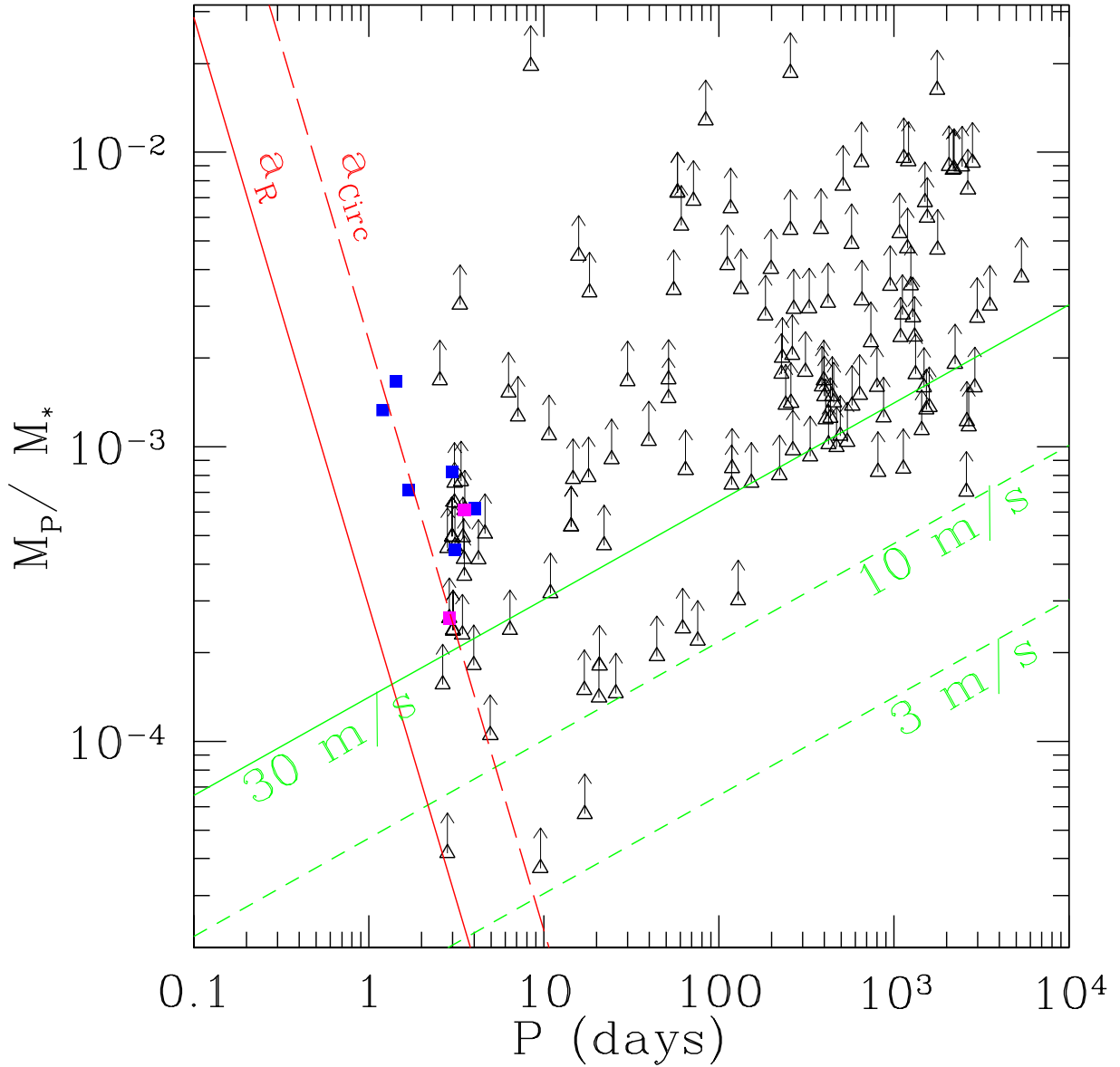}
\caption[fig1]{
%
%
Minimum mass ratio vs orbital period for the current observed
sample.  Planets discovered by radial velocity
surveys are shown as triangles with arrows indicating $1-\sigma$
uncertainties in mass due to unknown inclination.  The magenta squares
were discovered by radial velocity observations and have inclinations
and radii measured via transits.  The blue squares show planets
discovered by transit searches.  The green lines show the minimum mass
corresponding to various velocity semi-amplitudes and roughly indicate
where radial velocity surveys are nearly complete ($\ge30\,$m/s), have
significant sensitivity ($\ge10\,$m/s), and are only beginning to detect
planets ($\ge3\,$m/s).  The two red lines show the location of the
Roche limit ($a_R$) and the ideal circularization radius ($a_{\rm
circ}$) for a planet with a radius $R_P = 1.2\,R_J$.  The red
lines do not apply to the lowest mass planets that likely have a
radius significantly less than $1.2\,R_J$ given their
different internal structure. \\
\label{Fig1}}
\end{figure}


\begin{figure}[ht]
\plotone{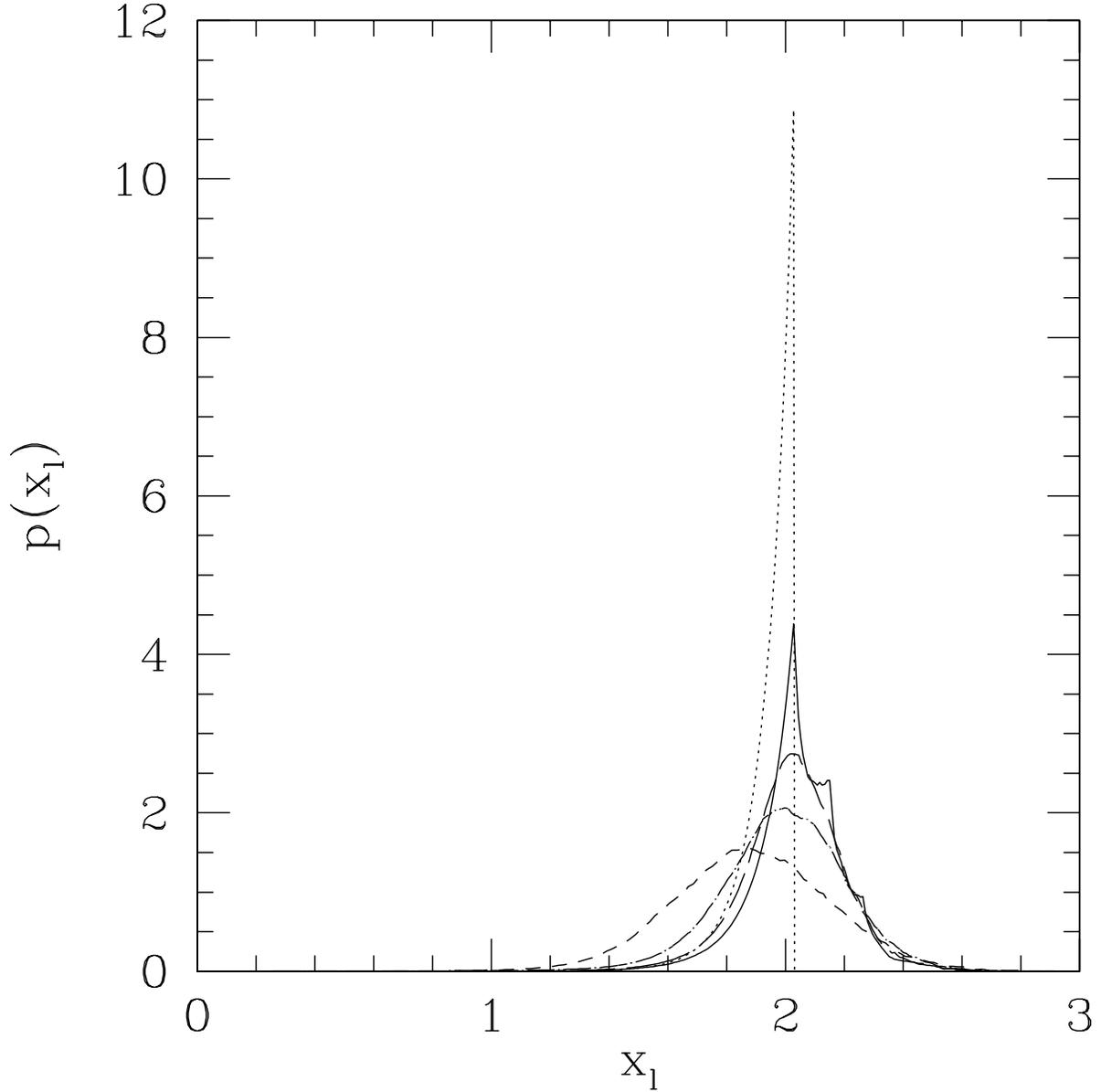}
\caption[fig2]{
%
%
Marginalized posterior probability distribution for $x_l$, the
lower cutoff for the ratio of a planet's semimajor axis
to the Roche limit.  Here we show multiple posterior distributions for
various simplified models.  The dotted curve assumes $R_P = 1.2\,R_J$
and $\sin i=1$ for all planets, while the remaining lines assume the
observed inclinations and radii for transiting planets and an
isotropic distribution of inclinations and mean radius $R_P =
1.2\,R_J$ for the remaining planets.  The widths of the distributions in
radii are $0.0 R_J$ (dotted and solid curves), $0.05 R_J$ (long dashes),
$0.1 R_J$ dots-dashes), $0.2 R_J$ (short dashes).  \\
\label{Fig2}}
\end{figure}


%
\begin{figure}[ht]
\plotone{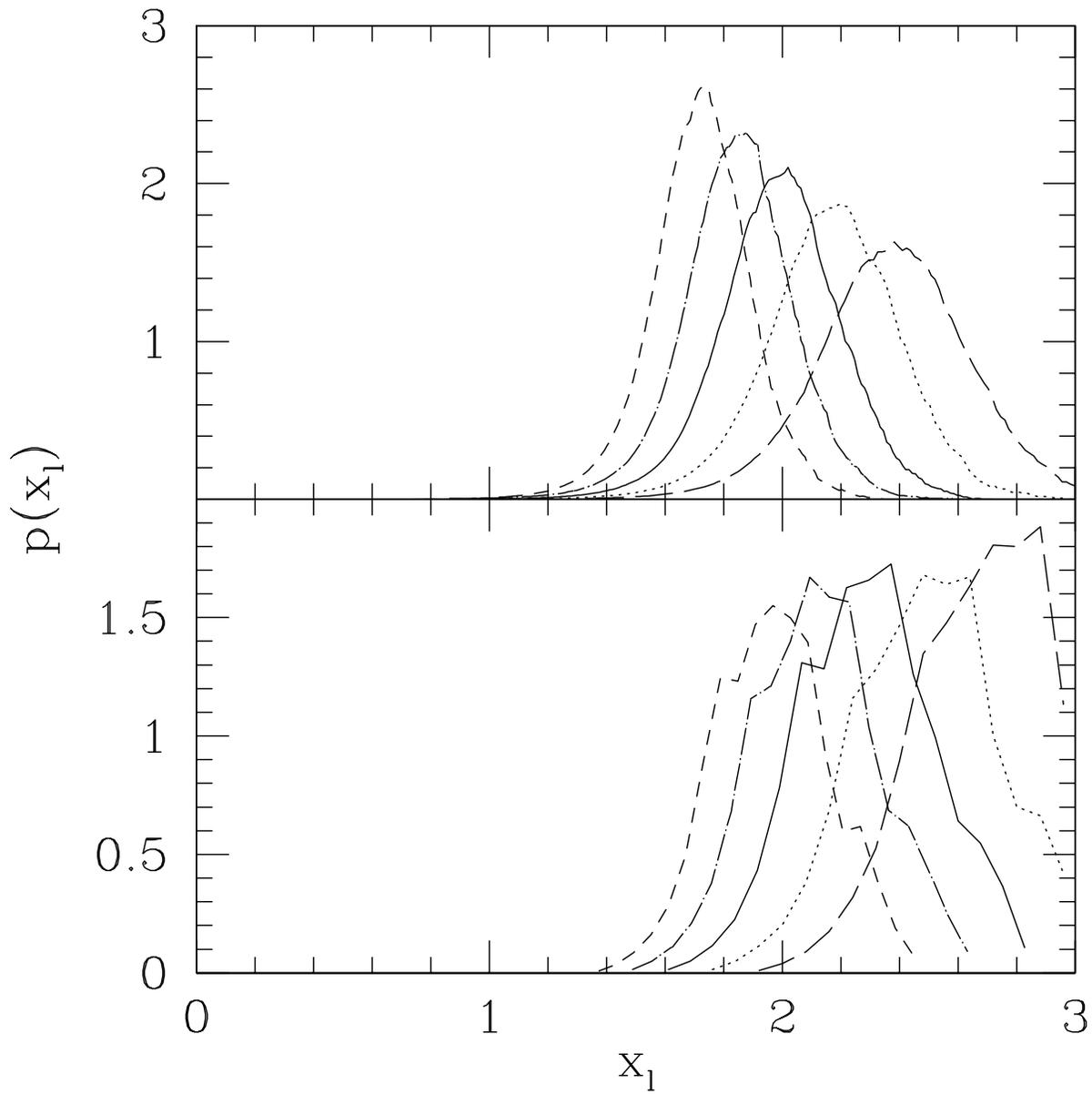}
\caption[fig3]{
%
%
Posterior distributions from our 1-d model (top) and full 2-d model  
(bottom) for
various mean radii: $\left<R_P\right> = 1.0
R_J$ (long dashes), $1.1 R_J$ (dotted), $1.2 R_J$ (solid), $1.3 R_J$
(dotted dashed), and $1.4 R_J$ (short dashes), all assuming $\sigma_ 
{R_P}
= 0.1 R_J$.   \\
\label{Fig3}}
\end{figure}


\end{document}